\documentclass[11pt]{article}

\usepackage{amssymb}
\usepackage{amsthm}
\usepackage{amsmath}

\usepackage{mathrsfs}  

\usepackage{graphicx} 
\usepackage{psfrag}
\usepackage{subfigure}

\usepackage{fancyheadings}

\graphicspath{{.}{/afs/aei-potsdam.mpg.de/u/jmh/figures/drawn_by_xfig/}}

\renewcommand{\sectionmark}[1]%
        {\markboth%
                {}%
                {{\rm\thesection}\quad{\sc #1}}}

\newcommand{\proofend}{\hspace*{\fill}\rule{0.2cm}{0.2cm}}

\newcommand{\textfrac}[2]{{\textstyle{\frac{#1}{#2}}}}

\newcommand{\tildenabla}{\tilde{\nabla}}

\newcommand{\weg}{\:\,}

\newcommand{\tilx}{\tilde{x}}
\newcommand{\tilt}{\tilde{t}}

\newcommand{\tfR}{{}^{\text{\tiny tf}}\!R}
\newcommand{\tfP}{{}^{\text{\tiny tf}}\!P}

\newcommand{\tfhR}{{}^{\text{\tiny tf}}\!\hat{R}}

\newcommand{\g}[3]{g^{\text{\tiny(#3)}}_{#1 #2}}
\newcommand{\hg}[3]{\hat{g}^{\text{\tiny(#3)}}_{#1 #2}}
\renewcommand{\k}[3]{k^{#1}_{\:\,#2 \text{\tiny (#3)}}}
\newcommand{\hk}[3]{\hat{k}^{#1}_{\:\,#2 \text{\tiny (#3)}}}
\newcommand{\sigmai}[3]{\sigma^{#1}_{\:\,#2 \text{\tiny (#3)}}}
\newcommand{\hsigma}[3]{\hat{\sigma}^{#1}_{\:\,#2 \text{\tiny (#3)}}}
\newcommand{\trk}[1]{\mathrm{tr}k_{\text{\tiny(#1)}}}
\newcommand{\trhk}[1]{\mathrm{tr}\hat{k}_{\text{\tiny(#1)}}}
\newcommand{\phii}[1]{\phi_{\text{\tiny(#1)}}}

\newcommand{\varkappai}[1]{\varkappa_{\text{\tiny(#1)}}}

\newcounter{store}
\newcounter{save}

\theoremstyle{plain}
\newtheorem{theorem}{Theorem}

\newtheorem{lemma}[theorem]{Lemma}

\theoremstyle{remark}

\begin{document}

\title{\huge \sc Power-law Inflation in Spacetimes without Symmetry}

\author{ \\
{\Large\sc J.\ Mark Heinzle}\thanks{Electronic address:  {\tt Mark.Heinzle@aei.mpg.de}} \\[1ex]
{\small \&} \\[1ex]
{\Large\sc Alan D.\ Rendall}\thanks{Electronic address:  {\tt Alan.Rendall@aei.mpg.de}} \\[2ex] 
Max Planck Institute for Gravitational Physics, \\
Albert Einstein Institute, \\
Am M\"uhlenberg 1, D-14476 Golm, Germany \\[2ex] }

\date{}
\maketitle
\begin{abstract}
We consider models of accelerated
cosmological expansion described by the Einstein equations coupled to a nonlinear
scalar field with a suitable exponential potential.
We show that homogeneous and isotropic solutions are stable
under small nonlinear perturbations without any symmetry assumptions.
Our proof is based on results on the nonlinear stability of de Sitter spacetime and
Kaluza-Klein reduction techniques.
\end{abstract}
\begin{center}

\begin{minipage}{12cm}
\begin{tabbing} Keywords: \= accelerated expansion, Kaluza-Klein reduction \kill
\> accelerated expansion, Kaluza-Klein reduction
\end{tabbing}
\end{minipage}

\end{center}

\vfill
\newpage

\section{Introduction}
\label{introduction}

At present the subject of accelerated cosmological expansion is of great
astrophysical interest. Many candidates have been suggested for the cause
of the acceleration, known under the name of dark energy. The greater part of the literature on this
concerns homogeneous (and even isotropic) models or their linearized perturbations. This is often enough to make
contact with observations. Nevertheless, since many of the phenomena of interest are linked to
inhomogeneities, it is desirable to develop an understanding of the full nonlinear dynamics for initial
data which are as general as possible. One
approach, which is pursued in this paper, is to try to prove general
mathematical theorems. A recent review of this approach is~\cite{Rendall:2004b}.

A well-known feature of models for accelerated cosmological expansion is
that they exhibit attractor solutions which are homogeneous and isotropic.
The simplest model is a positive cosmological constant in which case the
attractor is the de Sitter solution. In that case a theorem on stability
of this solution under small nonlinear perturbations has been proved~\cite{Friedrich:1986a}. It concerns the
vacuum equations with positive cosmological constant. No symmetry assumption is required. Generalizations
where matter such as a perfect fluid or kinetic theory is added are not available. Under the restriction of
plane symmetry an analogous result with collisionless matter was proved in~\cite{Tchapnda/Rendall:2003}. There
are also no generalizations of the result of~\cite{Friedrich:1986a} in the
literature to other models of accelerated expansion such as nonlinear scalar fields. This paper proves a
generalization of this type for a
nonlinear scalar field with exponential potential. Because it was
originally applied in models of the early universe this is associated with the name power-law inflation.
The method of proof restricts the exponents allowed to a discrete set.

The result of~\cite{Friedrich:1986a} makes essential use of the conformal properties of the Einstein equations
in four dimensions. Anderson~\cite{Anderson:2005}
has extended this analysis to to any even spacetime
dimension. Here we combine the results of~\cite{Anderson:2005} with a
Kaluza-Klein reduction which relates power-law inflation in four dimensions
with a suitable exponent in the potential to vacuum spacetimes with cosmological constant in higher
dimensions. It is proved that certain homogeneous and isotropic solutions are nonlinearly stable.

Consider a spacetime $(M,\tilde{g},\varphi)$ that satisfies the Einstein equations
\begin{equation}\label{Einsteinfield}
\tilde{R}_{\mu\nu} - \textfrac{1}{2} \tilde{R} \tilde{g}_{\mu\nu} = \kappa^2 T_{\mu\nu}
\end{equation}
with nonlinear scalar field matter
\begin{equation}\label{scalarenergymomentum}
T_{\mu\nu} = \nabla_\mu \varphi \nabla_\nu \varphi - 
\left[ \frac{1}{2} \nabla^\delta \varphi \nabla_\delta \varphi + V(\varphi) \right] \tilde{g}_{\mu\nu}\:,
\end{equation}
where $V(\varphi)$ is the potential of the scalar field.
Using that the spacetime is four-dimensional, 
Eq.~(\ref{Einsteinfield}) and~(\ref{scalarenergymomentum}) can be condensed into
\begin{subequations}\label{Esf}
\begin{equation}\label{Einsteinscalar}
\tilde{R}_{\mu\nu} = \kappa^2 \,\big( \nabla_\mu \varphi \nabla_\nu \varphi + V(\varphi) \tilde{g}_{\mu\nu}\big) \:.
\end{equation}
The Bianchi identity implies the equation of motion for the scalar field
\begin{equation}\label{waveeq}
\Box \varphi = V^{\prime}(\varphi)\:.
\end{equation}
\end{subequations}

Power-law inflation refers to the case
\begin{equation}\label{exppot}
V(\varphi) = V_0 \exp\left[-\kappa \lambda\,\varphi\right]
\end{equation}
with constants $V_0$ and $\lambda$, where $\lambda < \sqrt{2}$. 
Models with a potential of this type are the subject of the following.

The most elementary power-law inflation models are the homogeneous and isotropic models~\cite{Halliwell:1987},
of which the simplest is
\begin{equation}\label{hompower}
\tilde{g} =
-d \tilde{t}^2 + \left( \frac{d H}{2}\right)^{2 + \frac{4}{d}}\:
\tilde{t}^{2 + \frac{4}{d}}\: \delta_{i j} \,d\tilde{x}^i d\tilde{x}^j\:,
\end{equation}
where $d$ is related to the exponent $\lambda$ in the potential via $\lambda^2 = 2 d (2+d)^{-1}$.
The main theorem of this paper establishes nonlinear stability of this model.

\begin{theorem}\label{nonlinstabthm}
Consider the exponential potential
\begin{equation}\label{dpot}
V(\varphi) = V_0 \exp\left(-\kappa \sqrt{2} \sqrt{\frac{d}{d+2}}\,\varphi\right) \qquad \mathrm{with}\quad d \in\mathbb{N}, \,d \:\,\mathrm{even}\:.
\end{equation}
On $T^3$ consider smooth Cauchy initial data for the Einstein scalar field equations~\eqref{Esf} with potential~\eqref{dpot}.
Let the initial data be close in the $\mathcal{C}^\infty$-topology 
to the homogeneous and isotropic initial data of the model~\eqref{hompower}.
Then the initial data evolves into a power-law inflation model $(M,\tilde{g},\varphi)$ which is geodesically complete to the
future, and there exists
a coordinate system $(\tilde{t}, \tilde{x}^i)$ that is global to the future,
such that $\tilde{g}$ takes the form $\tilde{g} = -\alpha^2 d\tilde{t}^2 + \tilde{g}_{i j} d\tilx^i d\tilx^j$,
where $\tilde{g}_{i j}$ and $\varphi$ admit the asymptotic expansions
\begin{equation}\label{plasy}
\tilde{g}_{i j} = \tilde{t}^{2+4/d} \sum\limits_{m\geq 0} \tilde{g}_{i j}^{\text{\tiny\rm (m)}}\, \tilde{t}^{-2 m/d}\:,\quad
\varphi =  \kappa^{-1} \sqrt{2}\sqrt{\frac{d+2}{d}} \log \tilde{t} + \sum\limits_{m\geq 0} \varphi_{\text{\tiny\rm (m)}}\, \tilde{t}^{-2 m/d}\:,
\end{equation}
and $\alpha = \sum_{m \geq 0} \alpha_{\text{\tiny\rm (m)}}\, \tilde{t}^{-2 m/d}$.
Thus, the homogeneous and isotropic power-law inflation model \eqref{hompower} is nonlinearly stable.
\end{theorem}

A slightly different formulation of the theorem uses the concept of asymptotic simplicity.
We call the spacetime $(M, \tilde{g}, \varphi)$ asymptotically simple (in the future), 
when $\tilde{g}$ is conformal to a metric $\acute{g} = \Omega^{2+d} \tilde{g}$, and 
$\acute{\varphi} = \varphi + d \kappa^{-1} \lambda^{-1} \log \Omega$, where $\acute{g}$, $\acute{\varphi}$,
and the positive function $\Omega$ can be smoothly extended through the hypersurface $\Omega=0$,
which we denote as $\mathscr{I}^+$, the conformal boundary of $M$. 
The formula for the conformal transformation of the curvature scalar implies 
$\acute{g}^{\mu\nu} \nabla_\mu \Omega \nabla_\nu \Omega = -2\kappa^{-2} V_0 (2+d)^{-1} (3+d)^{-1} \exp(-\kappa\lambda \acute{\varphi})$ 
on $\mathscr{I}^+$, hence $\mathscr{I}^+$ is spacelike.
Note in this context that our definition for asymptotic simplicity differs from the standard
definition which applies for the vacuum case and for the case when matter can be neglected in an appropriate sense
in the neighborhood of the conformal boundary.

Theorem~\ref{nonlinstabthm} states that initial data close to homogeneous and isotropic data
evolves into an asymptotically simple solution.
We also have

\begin{theorem}\label{wellposedthm}
The initial value problem for the Einstein equations with scalar field matter and potential~\eqref{dpot},
where initial data is prescribed at conformal infinity $\mathscr{I}^+$, is well-posed.
The resulting spacetime $(M,\tilde{g},\varphi)$ is asymptotically simple with conformal boundary $\mathscr{I}^+$
and the metric and the scalar field admit the asymptotic expansions~(\ref{plasy}), where
the coefficients are uniquely determined by the initial data.
\end{theorem}

The proof of the theorems is based on the fact that 
power-law inflation models $(M, \tilde{g},\varphi)$ with potential~\eqref{dpot}
are in one-to-one correspondence with a certain type of $d$-dimensional
vacuum solutions $(\hat{M},\hat{g})$ of the Einstein equations with positive cosmological
constant. This is proved in Section~\ref{reduction}.
In Section~\ref{asyseries} we give a brief overview of 
the asymptotic behavior of solutions $(\hat{M},\hat{g})$; in particular
we recapitulate existence and nonlinear stability of asymptotically simple
solutions. On the basis of these results, 
the theorems are proved in Section~\ref{reductionseries}. 

Theorem~\ref{nonlinstabthm} formulates the asymptotic behavior in a certain coordinate system that
is not Gaussian in general. In Appendix~\ref{Gausscoords} we briefly discuss asymptotic expansions in Gaussian coordinates; in particular
we show that this choice of coordinates introduces logarithmic terms into the expansions.
In Appendix~\ref{formal} we investigate asymptotic expansions
for general exponential potentials on the level of formal power series. 
We demonstrate that logarithmic terms that appear in the series
can often be removed by a suitable choice of (non-Gaussian) coordinates;
this complements previous studies of
formal expansions of this type~\cite{Muller/Schmidt/Starobinsky:1990}.

\section{Reduction}
\label{reduction}

In this section we employ the Kaluza-Klein reduction method to find a one-to-one relationship
between solutions of the Einstein equations with positive cosmological constant $\Lambda$
and certain power-law inflation models. The formulation of Kaluza-Klein theory used and the notation 
is primarily based on~\cite{Choquet-Bruhat:1989}.

Consider a principal fiber bundle $G\rightarrow \hat{M} \stackrel{\pi}{\rightarrow} M$, 
where the base space $M$ is a 4-dimensional differentiable manifold,
and $G$ a $d$-dimensional Lie group (which we will eventually assume to be abelian).
On $\hat{M}$, let $\hat{g}$ be a Lorentzian metric such that
$\hat{g}$ is invariant under the right action of $G$ and
vertical vectors are not null w.r.t.\ $\hat{g}$.
The metric $\hat{g}$ induces a Lorentzian metric $g$ on $M$, a metric $\xi$ on each fiber, 
where $\xi$ is invariant
under the action of $G$, and a connection on $\hat{M}$ (in the form of a horizontal bundle).
In the so-called polarized case the horizontal distribution is assumed to be involutive and
in an adapted local trivialization over a chart neighborhood of $M$ with coordinates $\{x^\mu\,|\,\mu=0\ldots 3\}$ 
the metric $\hat{g}$ can be written as
\begin{equation}
\hat{g} = \hat{g}_{A B}\: e^A e^B = g_{\mu\nu}\, d x^\mu d x^\nu + \xi_{m n} \,\theta^m \theta^n\:,
\end{equation}
where the $\theta^m$ constitute a basis of right-invariant 1-forms on $G$,
and $g_{\mu\nu}$, $\xi_{m n}$ depend only on the coordinates $\{x^\delta\}$.
Greek indices run from $0$ to $3$, latin indices $m$, $n$, etc.\ assume values $1,2, \ldots, d$.
Capital letters $A$, $B$, etc.\ run over the combined range: the components $T_{A B}$ of a tensor
thus comprise $T_{\mu\nu}$, $T_{\mu n}$, $T_{m \nu}$, $T_{m n}$.

We set 
\begin{equation}
e^\phi := \sqrt{\det \xi} \qquad\qquad \text{so that} \qquad\qquad \xi_{m n} = e^{ 2 \phi/d}\, \zeta_{m n}\:,
\end{equation}
where $\det \zeta = 1$. Like $\xi_{m n}$, in the given trivialization, 
the field $\phi$ can be regarded as a scalar field on $M$.

To compute the Ricci tensor $\hat{R}_{A B}$ of the metric $\hat{g}$
we recall the general Kaluza-Klein formulas from~\cite{Choquet-Bruhat:1989}
and use $\nabla_\mu \phi = \frac{1}{2} \xi^{m n} \nabla_\mu \xi_{m n}$, where
$\xi^{m n}$ is the inverse of $\xi_{m n}$.
We obtain
\begin{subequations}\label{Riccihatone}
\begin{align}
\hat{R}_{\mu\nu} & = R_{\mu\nu} - \nabla_\mu \nabla_\nu \phi + \textfrac{1}{4} \nabla_\mu \xi_{m n} \nabla_\nu \xi^{m n} \\
\hat{R}_{m n} & = R_{m n} + \textfrac{1}{2} \xi^{p q} \nabla_\alpha \xi_{m p} \nabla^\alpha \xi_{n q} - 
\textfrac{1}{2} \left(\nabla^\alpha \nabla_\alpha + \nabla^\alpha \phi \,\nabla_\alpha \right) \xi_{m n}\:,
\end{align}
\end{subequations}
where $R_{\mu\nu}$ is the Ricci tensor of $g_{\mu\nu}$ and $R_{m n}$ the Ricci curvature of the fibers.
Both $R_{m n}$ and the components $\hat{R}_{\mu n}$ of the Ricci tensor $\hat{R}_{A B}$ vanish if the Lie group $G$ is abelian.

On $M$ we introduce the conformally rescaled metric
\begin{equation}
\tilde{g}_{\mu\nu} \::=\: e^{\phi}\: g_{\mu\nu} 
\end{equation}
and denote the Ricci curvature of $\tilde{g}_{\mu\nu}$ by $\tilde{R}_{\mu\nu}$.
Employing $\nabla^\alpha \nabla_\alpha + \nabla^\alpha \phi\, \nabla_\alpha = e^\phi\, \tilde{\nabla}_\alpha \tilde{\nabla}^\alpha$,
where $\tilde{\nabla}^\alpha = \tilde{g}^{\alpha\beta} \tilde{\nabla}_\beta$,
Eq.~(\ref{Riccihatone}) becomes
\begin{subequations}\label{Riccihattwo}
\begin{align}
\label{Riccihattwo1}
\hat{R}_{\mu\nu} & = \tilde{R}_{\mu\nu} + \frac{1}{2} \left(\tildenabla_\delta \tildenabla^\delta \phi\right) \tilde{g}_{\mu\nu} -
\left(\frac{1}{2} + \frac{1}{d}\right) \tildenabla_\mu \phi \tildenabla_\nu \phi  + 
\frac{1}{4} \tildenabla_\mu \zeta_{m n} \tildenabla_\nu \zeta^{m n} \\
\label{Riccihattwo2}
\hat{R}_{m n} & = R_{m n} - \frac{1}{2} e^\phi \,
\left[ \frac{2}{d} (\tilde{\nabla}_\alpha \tildenabla^\alpha \phi) \xi_{m n} +
e^{2 \phi/d} \left( \tildenabla_\alpha \tildenabla^\alpha \zeta_{m n} - 
\zeta^{p q} \tildenabla_\alpha \zeta_{m p} \tildenabla^\alpha\zeta_{n q}\right)
\right] 
\end{align}
\end{subequations}
Contracting~(\ref{Riccihattwo2}) with $\xi^{m n}$ leads to 
\begin{equation}\label{Rhat}
\xi^{m n} \hat{R}_{m n}  = \xi^{m n} R_{m n} - e^\phi\, \tilde{\nabla}_\alpha \tildenabla^\alpha \phi\:,
\end{equation}
where again $\xi^{m n} R_{m n} = 0$ in the case of an abelian Lie group $G$.
Eq.~(\ref{Riccihattwo}) simplifies when $\zeta_{m n}$ is independent
of $x^\delta$, i.e., when $\nabla_\mu \zeta_{m n} = 0$.

Assume that $(\hat{M},\hat{g})$ satisfies 
the Einstein vacuum equations with cosmological constant $\Lambda$, i.e.,
\begin{equation}\label{RicciLambda}
\hat{R}_{A B} = \frac{2}{d+2} \Lambda\: \hat{g}_{A B}\:.
\end{equation}
Suppose further that the Lie group $G$ is abelian and that $\nabla_\mu \zeta_{m n} = 0$.
From~(\ref{Rhat}) we obtain 
\begin{equation}\label{phieq}
\tilde{\nabla}_\alpha \tildenabla^\alpha \phi = - \frac{2 d}{d + 2}\, \Lambda \: e^{-\phi} \:.
\end{equation}
Eq.~(\ref{Riccihattwo1}) then leads to
\begin{equation}\label{Ricciinphi}
\tilde{R}_{\mu\nu} = \Lambda e^{-\phi}\, \tilde{g}_{\mu\nu} + 
\left(\frac{1}{2} +\frac{1}{d}\right) \tildenabla_\mu \phi \tildenabla_\nu \phi\:.
\end{equation}
Define
\begin{equation}\label{varphidef}
\varphi = \kappa^{-1} \sqrt{\frac{1}{2} +\frac{1}{d}}\; \phi\:,
\end{equation}
then~(\ref{phieq}) and~(\ref{Ricciinphi}) become
\begin{equation}\label{varphieq}
\tilde{\nabla}_\alpha \tildenabla^\alpha \varphi = V^{\prime}(\varphi) \qquad\text{and}\qquad
\tilde{R}_{\mu\nu} = \kappa^2  V(\varphi) \, g_{\mu\nu} + \kappa^2\, \tildenabla_\mu \varphi \tildenabla_\nu \varphi
\end{equation}
where
\begin{equation}\label{potentialdef}
V(\varphi) := \Lambda \kappa^{-2} \exp\left[-\kappa \sqrt{2} \:\sqrt{\frac{d}{d+2}}\;\varphi\right]\:.
\end{equation}
By comparing~(\ref{varphieq}) with~(\ref{Einsteinscalar}) and~(\ref{waveeq})
we conclude that $(M,\tilde{g},\varphi)$ is a solution of the Einstein equations
with nonlinear scalar field, where the potential is an exponential function,
i.e., $(M,\tilde{g},\varphi)$ is a power-law inflation model.
The exponent in the potential~(\ref{potentialdef}) is
\begin{equation}
\lambda = \lambda_d := \sqrt{2} \:\sqrt{\frac{d}{d+2}} < \sqrt{2}\:,
\end{equation}
cf.~(\ref{exppot}).

Conversely, given a solution $(M,\tilde{g},\varphi)$ representing power-law inflation
with exponent $\lambda= \lambda_d = (2 d)^{1/2} (d+2)^{-1/2}$ for some $d\in\mathbb{N}$,
we are able to construct a $(4+d)$-dimensional solution $(\hat{M},\hat{g})$ of 
the Einstein vacuum equations with positive cosmolog\nolinebreak ical constant. 
We take for $G$ an abelian Lie group, which ensures $\hat{R}_{\mu n} = 0$,
and we set $\partial_\mu \zeta_{m n}= 0$; e.g.,
we use $\hat{M} = M \times G$ where $G = \mathbb{R}^d$ is endowed with $\zeta_{m n} = \delta_{m n}$.
The relations~(\ref{varphieq}) together with the form of the potential
imply $\hat{R}_{A B} = 2/(d+2) \,\Lambda\: \hat{g}_{A B}$ with $\Lambda = \kappa^2 V_0$.

We conclude by giving a schematic overview of the one-to-one correspondence of solutions, which has been established:
\begin{equation}\label{corres}
\begin{array}{ccc}
(\hat{M}= M\times \mathbb{R}^d ,\hat{g}) & & (M,\tilde{g},\varphi) \\[2ex]
\hat{R}_{A B} - \frac{1}{2} \hat{R} \hat{g}_{A B} + \Lambda \hat{g}_{A B} = 0 & & 
\begin{array}{c}
\tilde{R}_{\mu\nu}  - \frac{1}{2} \tilde{R} \tilde{g}_{\mu\nu} = \kappa^2  T_{\mu\nu}[V(\varphi)]\:,\;
\tilde{\Box} \varphi = V^{\prime}(\varphi) \\[1ex]
V(\varphi) = \Lambda \kappa^{-2} \exp\left[-\kappa \sqrt{2} \:\sqrt{\frac{d}{d+2}}\;\varphi\right]
\end{array}
\\[4ex]
\hat{g} = g_{\mu\nu}\, d x^\mu d x^\nu + e^{2 \phi/d} \delta_{m n} \, d y^m d y^n 
& \quad & 
\tilde{g}_{\mu\nu} \,= \, e^{\phi} \:g_{\mu\nu} \:,\; \varphi = \kappa^{-1} \sqrt{\frac{d+2}{2d}}\, \phi 
\end{array}
\end{equation}

For the de Sitter solution in $(4+d)$ dimensions we can write
\begin{equation}
\hat{g} = -(d x^0)^2 + e^{2 H x^0} \left[ (d x^1)^2 + (d x^2)^2 + (d x^3)^2 + (d y^1)^2 + \ldots + (d y^d)^2 \right]\;,
\end{equation}
where $H^{-2} = (d+2)(d+3)/(2\Lambda)$. 
From~(\ref{corres}) we infer that $\phi = d H x^0$, and $\tilde{g}$ becomes
\begin{equation}\label{hompowerprov}
\tilde{g}_{\mu\nu} d x^\mu d x^\nu = e^{d H x^0} 
\left( -(d x^0)^2 + e^{2 H x^0} \left[ (d x^1)^2 + (d x^2)^2 + (d x^3)^2 \right] \right)\:.
\end{equation}
By introducing new coordinates $\tilde{x}^\mu$ through $d \tilde{x}^0 = \exp(d H x^0/2) \,d x^0$ and
$\tilde{x}^i = x^i$ we obtain
\begin{equation}
\tilde{g}_{\mu\nu}  d \tilde{x}^\mu d \tilde{x}^\nu =
-(d \tilde{x}^0)^2 + \left( \frac{d H}{2}\right)^{2 + \frac{4}{d}}\:
\left( \tilx^0\right)^{2 + \frac{4}{d}}\: \delta_{i j} \,d\tilde{x}^i d\tilde{x}^j\:,
\end{equation}
i.e., a flat Robertson-Walker model for power-law inflation as in~\eqref{hompower}.

\section{Asymptotic series}
\label{asyseries}

Consider the Einstein vacuum equations with cosmological constant $\Lambda$ in $n+1$ dimensions, $n\geq 3$, $n$ odd.
The $n+1$ decomposition of the equations consists of the constraint equations and the evolution equations 
\begin{equation}\label{evoeq}
\partial_t \,\hat{g}_{a b} = -2 \hat{g}_{a c} \hat{k}^c_{\weg b}\:,\quad
\partial_t \, \hat{k}^a_{\weg b} = \hat{R}^a_{\weg b} + (\mathrm{tr} \hat{k})\hat{k}^a_{\weg b} -\frac{2 \Lambda}{n-1}\delta^a_{\weg b}\:,
\end{equation}
where we have used a vanishing shift vector and a lapse function set equal to one.
In~\cite{Rendall:2004a} it was proved that the equations admit power series of the following type as formal solutions,
\begin{subequations}\label{gkser}
\begin{align}
\hat{g}_{a b} & = e^{2 H t} \:\left(\hg{a}{b}{0} + e^{-2 H t} \,\hg{a}{b}{2} + e^{-3 H t}\, \hg{a}{b}{3} + \ldots \right) \\[0.5ex]
\hat{k}^{a}_{\weg b} & = -H \delta^a_{\weg b} + \sum_{m\geq 2} \hk{a}{b}{m} \,e^{-m H t}\:,
\end{align}
\end{subequations}
where $H = \sqrt{2 \Lambda/[n (n-1)]}$. The coefficients $\hk{a}{b}{m} = \hsigma{a}{b}{m} + n^{-1} \trhk{m} \delta^a_{\weg b}$ 
are obtained recursively through the relations
\begin{subequations}\label{recalgsys}
\begin{equation}
[n - m] H \hsigma{a}{b}{m} = \sum_{p =2}^{m-2} \hsigma{a}{b}{p} \trhk{m-p} \,+\, \tfhR^a_{\weg b\:\text{\tiny(m)}}\,,\quad
[2 n - m] H \trhk{m}  = \sum_{p =2}^{m-2} \trhk{p} \trhk{m-p} \, + \, \hat{R}_{\text{\tiny(m)}}\:,
\end{equation}
for $m\geq 2$, $m\neq n$, $m\neq 2 n$, which follows from~(\ref{evoeq}), and through
\begin{equation}\label{algHconstraint}
2 (n -1) H \trhk{m} =  \hat{R}_{\text{\tiny(m)}} + \sum_{p=2}^{m-2} \left[ - \hk{a}{b}{p} \hk{b}{a}{m-p} + \trhk{p} \trhk{m-p}\right]\:,
\end{equation}
\end{subequations}
for $m =2 n$, which follows from the Hamiltonian constraint. 
$\hk{a}{b}{m}$ vanishes for all odd $m < n$. 
The evolution equation $\partial_t \,\hat{g}_{a b} = -2 \hat{g}_{a c} \hat{k}^c_{\weg b}$ implies that
the coefficients $\hg{a}{b}{m}$ are determined by the coefficients $\hk{a}{b}{l}$, $l=0 \ldots m$, recursively;
in particular $\hg{a}{b}{m} = 0$ for all odd $m < n$.
The remaining unspecified coefficients $\hg{a}{b}{0}$ and $\hg{a}{b}{n}$ encode the free data,
\begin{equation}\label{hatAhatB}
\hg{a}{b}{0} = \hat{A}_{a b} \;,\quad \hg{a}{b}{n} = \hat{B}_{a b}\:,
\end{equation}
where $\hat{A}_{a b}$ is a Riemannian metric, $\hat{B}_{a b}$ a symmetric tensor that satisfies $\hat{A}^{a b} \hat{B}_{a b} =0$
and $\hat{\nabla}^a \hat{B}_{a b} = 0$, where $\hat{A}^{a b}$ is the inverse of $\hat{A}_{a b}$ and
$\hat{\nabla}_a$ refers to $\hat{A}_{a b}$.

Consider now a spacetime $(\hat{M}, \hat{g})$ that is asymptotically simple and de Sitter (in the future), 
see, e.g.,~\cite{Friedrich:1986a}. By definition,
$\hat{g}$ is then conformal to a metric $\check{g} = \Omega^2 \hat{g}$, where $\check{g}$
and the positive function $\Omega$ can be smoothly extended through the hypersurface $\Omega=0$,
which is often denoted as the conformal boundary $\hat{\mathscr{I}}^+$ of $\hat{M}$.
Since $\check{g}^{\mu\nu} \nabla_\mu \Omega \nabla_\nu \Omega = -2 \Lambda n^{-1}(n-1)^{-1} = -H^2$ on $\hat{\mathscr{I}}^+$,
which follows from the conformal transformation of the curvature scalar, the metric $\check{g}$ takes the 
form 
\begin{equation}
\check{g} = -H^{-2} \alpha^2 d\Omega^2 + \check{g}_{a b} d \check{z}^a d \check{z}^b\:,
\end{equation}
when $\Omega$ is used as the time coordinate and the  
$\check{z}^a$ are spatial coordinates that are constant along the curves orthogonal to slices $\Omega =\mathrm{const}$.
The function $\alpha$ depends on $\Omega$ and $\check{z}^a$; $\alpha =1$ on $\hat{\mathscr{I}}^+$.
Letting $\hat{t} = \exp (-H\Omega)$ the physical metric becomes 
$\hat{g} = -\alpha^2 d\hat{t}^2 + \hat{g}_{a b} d\check{z}^a d\check{z}^b$ with $\hat{g}_{a b} = \exp(2 H \hat{t}) \check{g}_{a b}$.
In~\cite[Sec.~4]{Rendall:2004a}, in dimension $n=3$, 
it is shown that this relation together with an analogous set of 
fall-off conditions for $\alpha$ and $\hat{k}^a_{\weg b}$ implies that Gauss coordinates
can be introduced in which the metric $\hat{g}_{a b}$ and the extrinsic curvature $\hat{k}^a_{\weg b}$ exhibit an asymptotic 
expansion of the form~(\ref{gkser}).
It is straightforward to apply the proof in~\cite{Rendall:2004a} to all odd dimensions.

The initial value problem for the Einstein equations with positive cosmological constant, 
where initial data is prescribed at conformal infinity $\hat{\mathscr{I}}^+$, is well-posed;
this has been shown in~\cite{Friedrich:1986a} in the case $n=3$. In particular,
given an arbitrary Riemannian metric $\hat{A}_{a b}$ on a (compact) manifold $\hat{\mathscr{I}}^+$
and a symmetric tensor $\hat{B}_{a b}$ that is tracefree and divergence-free, then there
exists a unique future asymptotically simple solution of Einstein's equations whose
conformal boundary is $\hat{\mathscr{I}}^+$. 
Global non-linear stability of asymptotic simplicity has been proved in~\cite{Friedrich:1986b}. 
Hence, the evolution of initial data sufficiently close to de Sitter data 
yields a spacetime that is globally close to de Sitter and in particular 
asymptotically simple in the past and in the future.
For our purposes it is most relevant that these statements have been
generalized recently to arbitrary odd (spatial) dimensions $n$, in particular,
even-dimensional de Sitter spacetime is (globally) non-linearly stable,
see~\cite{Anderson:2005}.

We conclude that any initial data close to de Sitter
evolves into an asymptotically simple solution having an asymptotic expansion
of the form given in~(\ref{gkser}) together with~(\ref{hatAhatB}), where
$\hat{A}_{a b}$ and $\hat{B}_{a b}$ correspond to the conformal initial data set.

\section{Reduction of asymptotic series}
\label{reductionseries}

Consider the $(4+d)$-dimensional manifold $\hat{M} = M\times \mathbb{R}^d\,$; let $d$ be even.
On $\hat{M}$ consider solutions of the Einstein vacuum equations with cosmological constant $\Lambda$ of the type~(\ref{corres}), 
\begin{equation}\label{ghattype}
\hat{g} = -d t^2 + 
\underbrace{g_{i j} d x^i d x^j + e^{2 \phi/d} \delta_{m n} d y^m d y^n}_{\text{\normalsize $\hat{g}_{a b}\, d z^a d z^b$}}\:.
\end{equation}
If $(\hat{M},\hat{g})$ is asymptotically simple, as is guaranteed for solutions sufficiently close to de Sitter,
$\hat{g}_{a b}$ exhibits the asymptotics~\eqref{gkser}.
It follows that
\begin{equation}\label{gphiexp}
g_{i j} = e^{2 H t} \:\left(\g{i}{j}{0} + \sum_{m\geq 2} \g{i}{j}{m} \,e^{-m H t} \right)\:,\quad
\phi = d H t + \phii{0} + \sum_{m\geq 2} \phii{m} \, e^{-m H t}\:.
\end{equation}

From the $3+d$ split of $\hat{k}^a_{\weg b}$,
\begin{equation}
\hat{k}^a_{\weg b} \,\frac{\partial}{\partial z^a} \otimes d z^b = k^i_{\weg j}\, \frac{\partial}{\partial x^i}\otimes d x^j + 
\varkappa \delta^m_{\weg n} \,\frac{\partial}{\partial y^m}\otimes d y^n\:,
\end{equation}
we obtain in an analogous manner
\begin{equation}
k^i_{\weg j} = -H \delta^i_{\weg j} + \sum_{m\geq 2} \k{i}{j}{m} \,e^{-m H t}\:,\quad
\varkappa = -H + \sum_{m\geq 2} \varkappai{m}\,e^{-m H t} \:.
\end{equation}
The evolution equation $\partial_t \,\hat{g}_{a b} = -2 \hat{g}_{a c} \hat{k}^c_{\weg b}$, cf.~\eqref{evoeq}, reduces to 
\begin{equation}\label{evol1reduc}
\partial_t \,g_{i k} = -2 g_{i j} k^j_{\weg k}\:,\quad
\partial_t \, \phi = -d \varkappa\:,
\end{equation}
hence $\g{i}{j}{m}$ ($m\geq 2$) is determined recursively from $\g{i}{j}{0}$ and $\k{i}{j}{l}$, $l=2\ldots m$, and
$\varkappai{m} = m H d^{-1} \phii{m}$.
Reduction of the recursive algebraic system~\eqref{recalgsys} yields
\setcounter{store}{\value{equation}}
\begin{subequations}
\label{redalg}
\begin{align}
\label{redalg1}
[3 + d - m] H \sigmai{i}{j}{m}  & = \tfP^i_{\weg j\,\text{\tiny(m)}} + \sum_{p =2}^{m-2} \sigmai{i}{j}{p} [\trk{m-p} + d \varkappai{m-p}] \\
\label{redalg2}
[(6+d) - m] H \trk{m} + 3 H d \varkappai{m} &= P_{\text{\tiny(m)}} + \sum_{p =2}^{m-2} [\trk{p} + d\varkappai{p}] \,\trk{m-p} \\
\label{redalg3}
d H \trk{m} + [(3+2 d) -m] H d \varkappai{m} &= \rho_{\text{\tiny(m)}} + \sum_{p =2}^{m-2} [\trk{p} + d\varkappai{p}]\, d \varkappai{m-p}\:,
\end{align}
\end{subequations}
where
\begin{equation}
P^i_{\weg j} =  R^i_{\weg j} - \frac{1}{d} \nabla^i \phi \nabla_j \phi -\nabla^i \nabla_j \phi = 
\sum_{m\geq 2}  P^i_{\weg j\,\text{\tiny(m)}}\, e^{-m H t}
\end{equation}
and $\rho = - \nabla^i \nabla_i \phi - \nabla^i \phi \nabla_i \phi$ with an analogous expansion. 
(For the reduction it is useful to employ the warped product structure of the metric,
see, e.g.,~\cite{Dobarro/Unal:2005}.)
The $m$\raisebox{0.5ex}{\scriptsize th} 
coefficient $P^i_{\weg j\,\text{\tiny(m)}}$ is determined by the coefficients $\g{i}{j}{l}$ and $\phii{l}$, with $l=0\ldots (m-2)$;
$\rho_{\text{\tiny(m)}}$ by $\phii{l}$, $l=0\ldots (m-2)$.
The determinant of the coefficient matrix of the l.h.s.\ of (\ref{redalg2},\ref{redalg3}) is $(m-[d+3])(m-2[d+3])$.
The system~(\ref{redalg}) thus determines $\sigmai{i}{j}{m}$, $\trk{m}$, $\varkappai{m}$
recursively except for $m=d+3$, $m=2(d+3)$.
In the case $m=2(d+3)$ the system~(\ref{redalg2},\ref{redalg3}) is complemented by the equation
\setcounter{save}{\value{equation}}
\setcounter{equation}{\value{store}}
\begin{subequations}\setcounter{equation}{3}
\begin{multline}
2 H (2 +d) \left(\trk{m} + d \varkappai{m}\right) = - \sum_{p =2}^{m-2} \sigmai{i}{j}{p} \sigmai{j}{i}{m-p} +
\frac{2}{3} \sum_{p =2}^{m-2} \trk{p} \trk{m-p} \: +  \\ 
+ \frac{d-1}{d} \sum_{p =2}^{m-2}  d\varkappai{p} d \varkappai{m-p}
+ 2  \sum_{p =2}^{m-2}   d\varkappai{p} \trk{m-p} + P_{\text{\tiny(m)}} + \rho_{\text{\tiny(m)}} \:,
\end{multline}
\end{subequations}\setcounter{equation}{\value{save}}%
which is the reduced version of the constraint equation~(\ref{algHconstraint}).

The $0$\raisebox{0.5ex}{\scriptsize th} and the $(3+d)$\raisebox{0.5ex}{\scriptsize th} coefficients 
are undetermined by the algebraic system, they represent the free data.
When we decompose $\hat{A}_{a b}$, $\hat{B}_{a b}$, cf.~\eqref{hatAhatB}, according to
\begin{equation}\label{ABdecomp}
\hat{A}_{a b} d z^a d z^b = A_{i j} d x^i d x^j + e^{2 A} \delta_{m n} d y^m d y^n\:, \quad
\hat{B}_{a b} d z^a d z^b = B_{i j} d x^i d x^j + B \delta_{m n} d y^m d y^n
\end{equation}
we obtain
\begin{equation}\label{redfreedata}
\g{i}{j}{0} = A_{i j} \:, \quad\: \phii{0} =  d A \:,\qquad
\g{i}{j}{3+d} = B_{i j} \:, \quad (e^{2 \phi/d})_{\text{\tiny(3+d)}} = B\:.
\end{equation}
Hereby the data must satisfy the following conditions:
\begin{equation}\label{datacondi}
A^{i j} B_{i j} + d B e^{-2 A} = 0\:, \quad
\nabla^i B_{i j} - d B e^{-2 A} \nabla_j A + d B_{i j} A^{i k} \nabla_k A = 0\:, 
\end{equation}
where $A^{i j}$ is the inverse of $A_{i j}$
and $\nabla_i$ refers to $A_{i j}$.

We now make use of the relation~\eqref{corres} to prove Theorems~\ref{nonlinstabthm} \& \ref{wellposedthm}:

An asymptotically simple solution $(\hat{M},\hat{g})$ of the type~\eqref{ghattype} uniquely corresponds
to an asymptotically simple 
solution $(M,\tilde{g},\varphi)$ representing power-law inflation,
\begin{equation}\label{physmetred}
\tilde{g} = e^\phi\, \left( -d t^2 + g_{i j} d x^i d x^j \right)\:,\:\quad 
\varphi = \kappa^{-1} \sqrt{\frac{d+2}{2d}}\, \phi\:.
\end{equation}
Thus, the asymptotic behavior of $\tilde{g}_{\mu\nu}$ and $\varphi$
is uniquely determined by studying the
(reduction of the) asymptotic expansions of $\hat{g}$.
From the above analysis we obtain that
the asymptotic behavior of $\tilde{g}_{\mu\nu}$ and $\varphi$ is given
by the asymptotic series~\eqref{gphiexp} of $g_{i j}$ and $\phi$.
The coefficients in these series are determined via~\eqref{evol1reduc} 
through the coefficients $\k{i}{j}{l}$ and $\varkappai{l}$,
which are in turn determined recursively by~\eqref{redalg}.
The remaining free data is specified as in~\eqref{redfreedata}.

Introducing a new time coordinate $\tilde{t}$ through $d \tilde{t} = \exp(d H t/2) \,d t$ and
setting $\tilde{x}^i = x^i$, the metric~(\ref{physmetred}) becomes
\begin{equation}\label{tildeg}
\tilde{g} = 
 \underbrace{-e^{(\phi-d H t)}}_{\text{\normalsize $-\alpha^2$}} d\tilt^2 + 
\underbrace{e^{(\phi-d H t)} \left(\frac{d H}{2}\right)^{2+ \frac{4}{d}} \, \tilt^{2+\frac{4}{d}}\, h_{i j}}_{\text{\normalsize $\tilde{g}_{i j}$}}\, 
d\tilx^i d\tilx^j \:,
\end{equation}
where
\begin{subequations}\label{psih}
\begin{align}
\alpha^2 & = \exp\left[
d A + \sum_{m\geq 2} \phii{m} \left(\frac{d H}{2}\right)^{-2 m/d} \, \tilt^{-2 m/d} \right]\:,\\
h_{i j} & = A_{i j} + \sum_{m\geq 2} \g{i}{j}{m}\left(\frac{d H}{2}\right)^{-2 m/d} \,\tilt^{-2 m/d}\:.
\end{align}
\end{subequations}

To show Theorem~\ref{wellposedthm} consider on the three-dimensional manifold $\mathscr{I}^+$ 
a Riemannian metric $A_{i j}$, a symmetric tensor $B_{i j}$,
and fields $A$, $B$, that satisfy the condition~\eqref{datacondi}.
Defining $\hat{A}_{a b}$, $\hat{B}_{a b}$ according to~\eqref{ABdecomp}
results in Cauchy data at conformal infinity $\hat{\mathscr{I}}^+ = \mathscr{I}^+ \times \mathbb{R}^d$
for the Einstein vacuum equations with positive cosmological constant in $(4+d)$ dimensions.
The well-posedness of the corresponding Cauchy problem has been established in~\cite{Anderson:2005}.
Reduction of this result yields Theorem~\ref{wellposedthm}.

It is straightforward to show from~\eqref{redalg} that~(\ref{tildeg}) 
together with~(\ref{psih}) coincides with the homogeneous and isotropic  
solution~(\ref{hompower})
when $A=1$, $A_{i j} = \delta_{i j}$, and $(e^{2 \phi/d})_{\text{\tiny(3+d)}}= B= 0$, $\g{i}{j}{3+d} = B_{i j} =0$.
Since the solution~\eqref{hompower} uniquely corresponds to the $(4+d)$-dimensional
de Sitter solution, nonlinear stability of the latter reduces to
nonlinear stability of the former, which shows Theorem~\ref{nonlinstabthm}. 
Equivalently, Theorem~\ref{wellposedthm}
can be applied directly to obtain the result.
Hence, the evolution of initial data sufficiently close to data characterizing 
the power-law inflation Robertson-Walker model yields a spacetime that is globally close to
that model and the spacetime possesses a 
metric $\tilde{g}$ and
a scalar field $\varphi$ of the form~\eqref{physmetred}, which exhibit the asymptotic expansion~\eqref{gphiexp}
in a future end of the spacetime.

\section{Conclusions}

In this paper it has been shown that in certain models of accelerated
cosmological expansion homogeneous and isotropic solutions are stable
under small nonlinear perturbations without any symmetry assumptions.
These results concern the Einstein equations coupled to a nonlinear
scalar field with a suitable exponential potential. They show that
some known results for spacetimes with positive cosmological
constant generalize to a situation where the acceleration of a
cosmological model is due to the effect of a nonlinear scalar field.

For cosmological applications it would be desirable to incorporate
a description of ordinary matter (galaxies and dark matter) into
the models. It is expected that the source of the cosmological
acceleration (dark energy, the cosmological constant or the scalar
field) will dominate the dynamics at late times, but this should
be proved rather than assumed. In this paper we were not able to
include ordinary matter but note that this has not yet even been done
for a perfect fluid or collisionless matter in the case of a
cosmological constant. It seems that in order to do this methods
will be needed which are more direct than those using conformal
invariance properties.

Another direction in which the results should be extended is to
nonlinear scalar fields with more general potentials. The case
of an exponential potential with a general value of the exponent
is discussed on the level of formal power series in Appendix B
of this paper. The observation that a judicious choice of time
coordinate can simplify the asymptotic expansions may be useful
for later work using other methods. A similar discussion for a potential
with a strictly positive lower bound is given in~\cite{Bieli:2005}.
For wider classes of potentials the only mathematical theorems concern
spatially homogeneous spacetimes of Bianchi types I-VIII, including
normal matter~\cite{Lee:2005, Rendall:2004c, Rendall:2005}.

Questions related to cosmic acceleration and dark energy play a key
role in modern cosmology. They deserve the attention of researchers
in mathematical relativity and we hope that this paper will
contribute to the development of this area of mathematical physics.

\begin{appendix}
\numberwithin{equation}{section}   

\section{Asymptotics in Gaussian coordinates}
\label{Gausscoords}

The coordinates in which the asymptotic expansions have been given above are not Gaussian;
in this section we investigate the asymptotic expansions in Gaussian coordinates.
We begin by showing that the spacetime admits Gauss coordinates that are global to the future.

The metric and extrinsic curvature functions satisfy the following estimates:
\begin{subequations}\label{esti}
\begin{gather}
|\tilde{g}_{i j}|  \leq C \tilt^{2 + 4/d}  \:,\quad 
|\tilde{g}^{i j}|   \leq C \tilt^{-2 -4/d}    \:,\quad 
|\tilde{\Gamma}^i_{j k}|  \leq C \\
|\alpha - e^{d A/2}|   \leq C \tilt^{-4/d}   \:,\quad 
|\partial_{\tilt} \alpha |   \leq C \tilt^{-1-4/d}   \:,\quad 
|\partial_i \alpha |   \leq C \\
|\tilde{\sigma}^i_{\weg j}|   \leq C\tilt^{-1-4/d}  \:,\quad    
|\mathrm{tr} \tilde{k}    + (3 + 6/d) e^{-d A/2}\, \tilt^{-1} |  \leq C \tilt^{-1-4/d}
\end{gather}
\end{subequations}

\begin{lemma}
Consider a metric $\tilde{g}_{\mu \nu} d \tilx^\mu d \tilx^\nu = -\alpha^2 d \tilt^2 + \tilde{g}_{i j} d \tilx^i d \tilx^j$,
cf.~(\ref{tildeg}),
which is given on a time interval $[T,\infty)$, and
assume that there exists $C\in \mathbb{R}$ such that the estimates~(\ref{esti}) hold.
Consider a hypersurface $\tilt = \tilt_0$ and Gaussian coordinates based on that hypersurface.
If $\tilt_0$ is sufficiently large, then the Gaussian coordinates extend globally to the
future.
\end{lemma}

\proof
Consider an affinely parametrized geodesic $\gamma(\tau)$ that is orthogonal to the hypersurface $\tilt=\tilt_0$.
By a slight abuse of notation we write $(\tilt,\tilx)(\tau)$ for $\gamma(\tau)$;
\begin{equation}
\tilt(\tau_0) = \tilt_0\:, \quad \tilx(\tau_0) = \tilx^i_0\:, \quad
\frac{d \tilt}{d\tau}(\tau_0) = \alpha^{-1}(\tau_0)\:, \quad \frac{d \tilx^i}{d \tau}(\tau_0) = 0\:.
\end{equation}
It is a solution of the geodesic equations
\begin{subequations}\label{geodesicequations}
\begin{align}
\label{geodesiceqI}
\frac{d^2 \tilt}{d \tau^2} + \alpha^{-1} \partial_{\tilt}\alpha \left(\frac{d \tilt}{d \tau}\right)^2 + 
2 \alpha^{-1} \partial_i \alpha \frac{d \tilx^i}{d\tau} \frac{d\tilt}{d\tau} + 
\alpha^{-1} \tilde{k}_{i j} \frac{d \tilx^i}{d\tau} \frac{d\tilx^j}{d\tau} & = 0 \\
\label{geodesiceqII}
\frac{d^2 \tilx^i}{d \tau^2} + \alpha \partial^i\alpha \left(\frac{d \tilt}{d \tau}\right)^2 -
\frac{2}{3} \alpha \mathrm{tr} \tilde{k} \frac{d \tilx^i}{d\tau} \frac{d\tilt}{d\tau} -
2\alpha \tilde{\sigma}^i_{\weg j} \frac{d \tilx^j}{d\tau} \frac{d\tilt}{d\tau} +
\tilde{\Gamma}^i_{j k}  \frac{d \tilx^i}{d\tau} \frac{d\tilx^j}{d\tau} & = 0\:.
\end{align}
\end{subequations}
Let $\epsilon>0$ be sufficiently small in comparison to $\alpha^{-1}(\tau_0)$ and $E> 0$, let $\zeta \in (1 + 2/d, 1 + 4/d)$,
and consider the maximal interval $[\tau_0, \bar{\tau})$ such that
\begin{equation}\label{firstass}
|\frac{d\tilt}{d\tau} - \alpha^{-1}| \leq \epsilon \:, \quad
\tilt^\zeta \,|\frac{d\tilx^i}{d\tau}| \leq E \quad \text{ on $[\tau_0,\bar{\tau})$}.
\end{equation}
Integrating~(\ref{firstass}) we infer that there exist constants $C_+ > C_- > 0$, such that
\begin{equation}\label{tilttau}
C_- (\tau -\tau_0) < \tilt(\tau) -\tilt_0 < C_+  (\tau -\tau_0)\:, \quad
|\tilx^i(\tau) -\tilx^i_0| < E C_-^{-1} (\zeta-1)^{-1}\, \tilt_0^{1-\zeta} 
\end{equation}
on $[\tau_0,\bar{\tau})$.

Making use of~(\ref{esti}) the geodesic equation~(\ref{geodesiceqI}) yields 
\begin{equation}
\frac{d^2 \tilt}{d \tau^2} = \iota(\tau) \quad\text{where}\quad |\iota(\tau)| \leq \mathrm{const}\, \tilt^{1-2 \zeta + 4/d}\;,
\end{equation}
and by integration
\begin{equation}\label{improve1}
|\frac{d\tilt}{d\tau}(\tau) - \alpha^{-1}(\tau_0)| \leq \mathrm{const}\: \tilt_0^{2(1+2/d-\zeta)}\:,\quad \text{hence}\quad
|\frac{d\tilt}{d\tau} - \alpha^{-1}| \leq \mathrm{const}\: \tilt_0^{2(1+2/d-\zeta)}\:,
\end{equation}
where the constants depend on $\epsilon$, $E$, but are independent of $\tilt_0$.
The geodesic equation~(\ref{geodesiceqII}) can be treated by noting that
\begin{equation}
\alpha \mathrm{tr}\tilde{k} = (\alpha- e^{d A/2}) \mathrm{tr}\tilde{k} + 
e^{d A/2} \left(\mathrm{tr}\tilde{k} + [3+\frac{6}{d}] e^{-d A/2} \tilt^{-1}\right) - [3+\frac{6}{d}]\, \tilt^{-1}\:.
\end{equation}
By~(\ref{esti}) we obtain
\begin{equation}
\frac{d^2 \tilx^i}{d \tau^2} + \left( 2 + \frac{4}{d}\right) \tilt^{-1} \frac{d\tilt}{d\tau} \frac{d\tilx^i}{d\tau} = 
\varsigma(\tau) \quad\text{where}\quad |\varsigma(\tau)| \leq \mathrm{const}\, \tilt^{-2 - 4/d}\;;
\end{equation}
integration leads to
\begin{equation}\label{improve2}
|\frac{d\tilx^i}{d\tau}| \leq \mathrm{const} \:\left(\tilt^{-1-4/d} - \tilt_0 \tilt^{-2 - 4/d}\right)
\quad\text{and}\quad
\tilt^\zeta \,|\frac{d\tilx^i}{d\tau}| \leq \mathrm{const} \:\tilt_0^{- 1 - 4/d + \zeta}\:.
\end{equation}
If $\tilt_0$ is sufficiently large, then $\mathrm{const}\: \tilt_0^{2(1+2/d-\zeta)} < \epsilon$
and $\mathrm{const} \:\tilt_0^{- 1 - 4/d + \zeta} < E$, hence~(\ref{improve1}) and~(\ref{improve2})
improve~(\ref{firstass}) on $[\tau_0,\bar{\tau})$. Since $\bar{\tau}$ was chosen maximal, $\bar{\tau}$ must be infinite,
and the above estimates hold globally. In particular, from~(\ref{improve2}),
\begin{equation}\label{timelikeasyprop}
|\frac{d\tilx^i}{d\tau}| \leq \mathrm{const} \:\tilt^{-1-4/d}
\quad\text{and}\quad
|\frac{d\tilt}{d\tau} - \alpha^{-1}| \leq \mathrm{const}\: \tilt^{-4/d}\:,
\end{equation}
where the second inequality results from the fact that the geodesic is affinely parametrized.
Global existence of the timelike geodesics orthogonal to the
hypersurface $\tilt=\tilt_0$ has thus been established; the asymptotic properties are given by~(\ref{timelikeasyprop}).

To show that the family of geodesics gives rise to a global Gaussian coordinate system we
investigate geodesic deviation.
Consider the geodesic $\gamma(\tau)$ and let $d_j^{\weg \mu}$ be the deviation vector 
between $\gamma$ and an infinitesimally neighboring geodesic in the direction $\partial_j$, i.e.,
\begin{equation}
d_j^{\weg 0}(\tau_0) = 0 \:, \quad d_j^{\weg i}(\tau_0) = \delta^i_j \:,\quad
\frac{d d_j^{\weg 0}}{d\tau}(\tau_0) = -(\alpha^{-2} \partial_j \alpha) (\tau_0)\:,\quad
\frac{d d_j^{\weg i}}{d\tau}(\tau_0) = 0\:.
\end{equation}
In analogy to~(\ref{firstass}) we can assume that $|d d_j^{\weg 0}/d\tau + \alpha^{-2} \partial_j\alpha|\leq \epsilon$
and $\tilt^\zeta |d d_j^{\weg i}/d\tau| \leq E$ on a maximal interval $[\tau_0,\bar{\tau})$.
Along the lines of the above argument, by using derivatives of the geodesic equations~(\ref{geodesicequations}) w.r.t.\ spatial variables
we can improve these inequalities to obtain $\bar{\tau} =\infty$, and we get that
\begin{equation}
|\frac{d d_j^{\weg i}}{d\tau} |  \leq \mathrm{const} \: \tilt^{-1-4/d} \quad\text{so that} \quad
|d_j^{\weg i} - \delta^i_j| \leq \mathrm{const} \: \tilt_0^{-4/d}
\end{equation}
on $[\tau_0,\infty)$.
For large $\tilt$ the component $d_j^{\weg 0}$ will increase linearly in $\tilt$. In line with the existing gauge
freedom we may redefine $d_j^{\weg \mu}$,
\begin{equation}
\tilde{d}_j^{\weg \mu} =
\begin{pmatrix}
d_j^{\weg 0} \\ d_j^{\weg i} 
\end{pmatrix}
+
\lambda(\tilt)
\begin{pmatrix}
d\tilt/ d\tau\\
d\tilx^i/ d\tau
\end{pmatrix}
=
\begin{pmatrix}
0 \\
\delta^i_j + \mathrm{const} \,[\tilt_0^{-4/d} +  \tilt^{-4/d}]
\end{pmatrix}
\end{equation}
for a suitable choice of $\lambda(\tilt)$.
We infer that the deviation vector behaves in a nice manner, at least for $\tilt_0$ sufficiently large,
so that the family of geodesics originating from $\tilt=\tilt_0$ generates a Gaussian
coordinate system that is global in the future.
\proofend

Let $\{\tau,\tilx^i\}$ denote the Gaussian coordinate system constructed above.
In these coordinates the field equations take the form $\partial_\tau \tilde{g}_{i j} = -2 \tilde{k}_{i j}$, 
\begin{subequations}
\begin{align}\label{gaufe}
& \partial_\tau \tilde{k}^i_{\weg j}  = \tilde{R}^i_{\weg j} + (\mathrm{tr} \tilde{k}) \tilde{k}^{i}_{\weg j} - 
8\pi S^i_{\weg j} + 4 \pi \delta^i_{\weg j} (\mathrm{tr} S -\rho)\:, \\
- & \,\partial_\tau^2 \varphi +  \tilde{\Delta} \varphi + (\mathrm{tr}\tilde{k}) \partial_\tau \varphi = V^\prime(\varphi)\:,
\end{align}
\end{subequations}
where $S_{i j}$ and $\rho$ stem from the scalar field energy-momentum tensor~(\ref{scalarenergymomentum}), i.e.,
$\rho = T_{00}$, $S_{i j} = T_{i j}$.

It can be shown by example that series of the type
\begin{equation}
\tilde{g}_{i j} = \tau^{2 + 4/d} \sum\limits_{m \geq 0} \g{i}{j}{m}  \tau^{-2 m/d}\:,\quad
\varphi = [2 \kappa^{-2} (d+2)/d]^{1/2} \log \tau + 
\sum\limits_{m \geq 0} \varphi_{\text{\tiny(m)}} \, \tau^{-2 m/d}
\end{equation}
do not provide solutions of~\eqref{gaufe} in general. 
The asymptotic expansions of $\tilde{g}$, $\varphi$, etc.\ necessarily include logarithmic terms, i.e.,
\begin{equation}\label{logser}
\tilde{g}_{i j}  = \tau^{2 + 4/d} \sum\limits_{m \geq 0} \sum\limits_{l = 0}^{L_m} \g{i}{j}{m,l} (\log \tau)^l\, \tau^{-2 m/d}\:, \,\text{etc.} \,,
\end{equation}
where $L_m \in \mathbb{N}\:\,\forall m$.
It turns out that $L_m = 0$ for all $m < d/2$ in the case $d = 4 k$, $k\in\mathbb{N}$, and
$L_m = 0$ for all $m< 3 +d$ in the case $d = 2 (2 k +1)$, $k \in \mathbb{N}$.

It is interesting to contrast~\eqref{logser} and~\eqref{plasy}:
in Gaussian coordinates the asymptotic expansions contain logarithmic terms in general,
however, by the use of a suitable time coordinate these logarithmic terms can be removed.
In Appendix~\ref{formal} we investigate whether this statement can be generalized, on the level of formal power series,
for arbitrary exponents $d$.

\section{Formal asymptotic expansions with general exponents}
\label{formal}

For a spacetime $(M,\tilde{g}, \varphi)$ 
consider the Einstein scalar field equations with potential
\begin{equation}\label{pot}
V(\varphi) = V_0 \exp\left(-\kappa \sqrt{2} \sqrt{\frac{d}{d+2}}\,\varphi\right)\qquad \text{with}\quad  d \in\mathbb{R}\,.
\end{equation}
In~\cite{Muller/Schmidt/Starobinsky:1990} it was shown that the equations admit power series as formal solutions.
The analysis was performed in Gaussian coordinates $\{\tau, \tilx^i\}$, so that $\tilde{g} = -d\tau^2 + \tilde{g}_{i j} d \tilx^i d\tilx^j$,
and it was found that
\begin{equation}
\tilde{g}_{i j} = \tau^{2 + 4/d} \sum\limits_{m \in M} \g{i}{j}{m}  \tau^{-2 m/d}\:,\quad
\varphi = [2 \kappa^{-2} (d+2)/d]^{1/2} \log \tau + 
\sum\limits_{m \in M} \varphi_{\text{\tiny(m)}} \, \tau^{-2 m/d}\:,
\end{equation}
where $M$ is the set $\{0\} \cup \{ (d/2) n_1 + 2 n_2 + (3+d) n_3 \:|\: n_i \in \mathbb{N}\}$.
However, the series contain logarithmic terms, cf.~\eqref{logser},
if there exists $n_1, n_2 \in \mathbb{N}$ such that $(d/2) n_1 + 2 n_2  = (3+d)$, which is the case
when $d =n$ or $d = 6/n$, $n\in\mathbb{N}$.

The problem simplifies considerably when we make the ansatz
\begin{equation}\label{physmetred2}
\tilde{g} = e^\phi\, \left( -d t^2 + g_{i j} d x^i d x^j \right)\:,\:\quad 
\varphi = \kappa^{-1} \sqrt{\frac{d+2}{2d}}\, \phi\:,
\end{equation}
which is inspired by~(\ref{corres}).
Let $k^i_{\weg j}$ denote the second fundamental form of the hypersurfaces $t=\mathrm{const}$ 
in the spacetime $(M, -dt^2 +g)$, and $\sigma^i_{\weg j}$ its trace-free part. 
Then the Einstein scalar field (evolution) equations become $\partial_t g_{i j} = -2 g_{i l} k^l_{\weg j}$ and
\begin{subequations}
\begin{align}
\partial_t \sigma^l_{\weg j} & = \tfR^l_{\weg j} + (\mathrm{tr} k) \sigma^l_{\weg j} - 
\frac{1}{d} \left( \nabla^l \phi \nabla_j \phi - \frac{1}{3} \nabla^k \phi \nabla_k \phi\right)
- \left( \nabla^l  \nabla_j \phi - \frac{1}{3} \nabla^k \nabla_k \phi\right) \\
\partial_t \mathrm{tr} k & = R +(\mathrm{tr} k)^2 - \kappa^2 V_0 \frac{2}{d+2} - \partial_t \phi \,\mathrm{tr} k - 
\frac{1}{d} \nabla^k \phi \nabla_k \phi - \nabla^k \nabla_k \phi \\
\Box \phi & + \nabla^k \phi \nabla_k \phi - (\partial_t \phi)^2 = -\kappa^2 \frac{2 d}{d+2} \, V_0\,,
\end{align}
\end{subequations}
where $R$ and $\nabla$ refer to $g$.
It can be proved that this system of equations, complemented by the constraints,
admits power series of the following type as formal solutions:
\begin{equation}\label{gphiexp2}
g_{i j} = e^{2 H t} \:\left(\g{i}{j}{0} + \sum_{m \in \mathcal{M}} \g{i}{j}{m} \,e^{-m H t} \right)\:,\quad
\phi = d H t + \phii{0} + \sum_{m \in \mathcal{M}} \phii{m} \, e^{-m H t}\:,
\end{equation}
cf.~\eqref{gphiexp},
where $H^2 = 2 V_0 \kappa^2 (d+2)^{-1}(d+3)^{-1}$ and 
$\mathcal{M} = \{ 2 m_1 + (3+d) m_2 \:|\: m_i \in \mathbb{N}\}$.
The recursive algebraic system specifying the coefficients is
essentially identical with~\eqref{redalg}, the free data is represented by the 
$0$\raisebox{0.5ex}{\scriptsize th} and the $(3+d)$\raisebox{0.5ex}{\scriptsize th} coefficients.
In this context the exponent $d \in \mathbb{R}$ is still not completely arbitrary, though:
for $d = 2k +1$, $k\in\mathbb{N}$, the expansions~(\ref{gphiexp2}) must be supplemented by logarithmic terms.

These results suggest that there exist two types of logarithmic terms in formal expansions:
(i) artificial logarithms which are due to an unsuitable choice of coordinates and can be removed
by using different coordinates, and
(ii) genuine logarithms.
In our particular case we have seen that, of the logarithmic terms in~\cite{Muller/Schmidt/Starobinsky:1990} 
which appear for $d =n$ and $d = 6/n$, $n\in\mathbb{N}$,
only those in the case $d = 2k +1$, $k\in\mathbb{N}$ can be genuine.

\end{appendix}


\bibliographystyle{plain}

\end{document}